\documentclass{tlp}
\nocite{*}

\usepackage{aopmath}

\usepackage{psfrag,graphicx,flafter}

\newcommand{\irule}[2]{\frac{\textstyle\rule[-1.3ex]{0cm}{3ex}#1}%
{\textstyle\rule[-.5ex]{0cm}{3ex}#2}}

\def\0{{\mathbf 0}}
\def\1{{\mathbf 1}}

\def\lbb{\mathopen{\{\kern-.25em|}}
\def\rbb{\mathclose{|\kern-.25em\}}}
\def\comp#1{\lbb#1\rbb}

\newcommand\kas{\mathop{\mathsf{Kas}}}
\newcommand\tgs{\mathop{\mathsf{Tgs}}}

\newcommand\Na{\mathit{Na}}

\newcommand\Nb{\mathit{Nb}}

\newcommand\ak{\mathit{authK}}

\newcommand\at{\mathit{authTicket}}
\newcommand\sk{\mathit{servK}}
\newcommand\st{\mathit{servTicket}}
\newcommand\Tt{\mathit{Ts}}
\newcommand\Tuno{\mathit{T_1}}
\newcommand\Tdue{\mathit{T_2}}
\newcommand\Ttre{\mathit{T_3}}
\newcommand\Ta{\mathit{Ta}}
\newcommand\Ts{\mathit{Ts}}

\newcommand\authenticatoruno{\mathit{authenticator1}}
\newcommand\authenticatordue{\mathit{authenticator2}}
\newcommand\authenticatortre{\mathit{authenticator3}}

\newcommand\Ka{\mathit{Ka}}
\newcommand\Kb{\mathit{Kb}}

\newcommand\Ktgs{\mathit{Ktgs}}

\newcommand\Kd{\mathit{Kd}}

\newcommand\Kab{\mathit{Kab}}

\def\ent{\vdash}
\newcommand\deff{\mathit{def}}

\newcommand\Deff{\mathit{Def}}
\def\defb{\mathopen{\deff\kern-.25em_B}}

\newcommand{\RefFig}[1]{Figure\nolinebreak\hspace{0.25em}\ref{#1}}

\newtheorem{definition}[theorem]{Definition}

\newif\ifremarks
\remarksfalse
\newcommand{\Commento}[1]{\ifremarks{\bf\footnotesize \$. #1}\fi}

\submitted{18 March 2002}
\revised{14 January 2003, 19 August 2003} 
\accepted{9 December 2003}

\begin{document}

\title{Soft Constraint Programming to Analysing Security Protocols}
\author[G. Bella and S. Bistarelli]
{GIAMPAOLO BELLA\\
University of Cambridge, Computer Laboratory\\
15 JJ Thomson Avenue, Cambridge CB3 0FD, UK\\
         \email{giampaolo.bella@cl.cam.ac.uk} \and
         \vspace{-0.7cm}\\
Universit\`a di Catania, Dipartimento di Matematica e Informatica\\
Viale A. Doria 6, I-95125 Catania, Italy\\
         \email{giamp@dmi.unict.it} \and
STEFANO BISTARELLI\\
CNR, Istituto di Informatica e Telematica\\
Via G. Moruzzi 1, I-56124 Pisa, Italy\\
         \email{stefano.bistarelli@iit.cnr.it}\and
         \vspace{-0.7cm}\\
Universit\`a ``D'Annunzio'', Dipartimento di Scienze\\
Viale Pindaro 42, I-65127 Pescara, Italy\\
         \email{bista@sci.unich.it}}
\pagerange{\pageref{firstpage}--\pageref{lastpage}}


\label{firstpage}
\maketitle

\begin{abstract}
Security protocols stipulate
how the remote principals of a computer network should interact in order
to obtain specific security goals.
The crucial goals of {\em confidentiality} and {\em authentication} may
be achieved in various forms, each of different strength.
Using soft (rather than crisp) constraints,
we develop a uniform formal notion for the two goals. They are
no longer formalised as mere yes/no properties as in the
existing literature, but gain an extra parameter, the {\em
security level}. For example, different messages can enjoy different levels
of confidentiality, or a principal can achieve different levels of
authentication with different principals.

The goals are formalised within a general framework for protocol analysis
that is amenable to mechanisation by model checking. Following the application
of the framework to analysing the asymmetric Needham-Schroeder
protocol~\cite{security-padl01,bella-bista-cambridge02}, we have recently
discovered a new attack on that protocol as a form of retaliation by principals
who have been attacked previously. Having commented on
that attack, we then demonstrate the framework on a bigger,
largely deployed protocol consisting of three phases, Kerberos.\\
\begin{center}
{\bf To appear in Theory and Practice of Logic Programming (TPLP)}
\end{center}
\end{abstract}
%


\section{Overview}
A number of applications ranging from electronic transactions over the
Internet to banking transactions over financial networks make use of
security protocols. It has been shown
that the protocols often fail to meet
their claimed goals~\cite{prudent,lowenew}, so a number of
approaches for analysing them formally
have been
developed~\cite{lowefirst,miojucs,paulsonjcs,degano1,degano2,gorrieri1,gorrieri2,abadi1,abadi2}.
The threats to the protocols come from malicious principals who
manage to monitor the network traffic building fake messages at
will. A major protocol goal is {\em confidentiality}, confirming
that a message remains undisclosed to malicious principals.
Another crucial goal is {\em authentication}, confirming a
principal's participation in a protocol session.
These goals are formalised
in a mere ``yes or no'' fashion in the existing literature. One
can just state whether a key is confidential or not, or whether a principal
authenticates himself with another or not.

\paragraph{Security goals are not simple boolean properties.}
``Security is not a simple boolean predicate; it concerns how well a system
performs certain functions''~\cite{andersonwhy}. Indeed, experience shows
that system security officers
exercise care in applying any firm boolean statements to the real world
even if they were formal.
In general, formal security proofs are conducted within simplified models.
Therefore, security officers attempt to bridge the gap between
those models and the real word by adopting the largest possible variety of
security measures all together.
For example, firewalls accompany SSH connections. Limiting the access to certain ports
of a server is both stated on the firewall {\em and} on the server itself.
Biometric technology recently set aside the use of passwords to strengthen
authentication levels of principals.
Still, principals' credentials can be constrained within a validity time interval.
The officer shall balance the cost of an extra security measure with
his perception of the unmanaged risks.
Any decision will only achieve a certain {\em security level}.

Security levels also characterise security patches \cite{simon-private}.
Each patch in fact comes with a
recommendation that is proportionate to the relevance of the security hole the patch is meant to fix.
Patches may be critical, or recommended, or suggested, or software upgrade, etc.
Depending on the cost of the patch and on the relevance of the
hole, the security officer can decide whether or not to upgrade the system.
It is a security policy what establishes the maximum level up until
a patch can be ignored.

This all confirms that real-world security is based on security levels rather than on categorical,
definitive, security assurances. In particular, security levels characterise the protocol
goals of confidentiality and authentication.
Focusing on the former goal, we remark that different messages
require ``specific degrees of protection against disclosure''~\cite{stefatrovatasuweb}. For example, a user password requires higher protection than
a {\em session key}, which is only used for a single protocol session.
Intuitively, a password ought to be ``more confidential'' than a session key.
Also, a confidentiality attack due to off-line cryptanalysis should not be
imputed to the protocol design.
Focusing on authentication, we observe that
a certificate stating that $K$ is a principal $A$'s
{\em public key} authenticates
$A$ very weakly. The certificate only signifies that $A$ is a registered
network principal, but in fact confers no guarantee about $A$'s participation
in a specific protocol session.
A message signed by $A$'s {\em private key} authenticates $A$ more strongly,
for it signifies that $A$ participated in the protocol in order to sign
the message.

\paragraph{Our original contributions.}
We have developed enriched formal notions for the two goals.
Our definitions of $l$-{\em confidentiality} and of $l$-{\em authentication} highlight
the security level $l$.
One of the advantages of formalising security levels
is to capture the real-world non-boolean concepts of confidentiality
and authentication.

Each principal assigns
his own security level to each message
--- different levels to different messages ---
expressing the principal's trust on the confidentiality of the message.
So, we can formalise that different goals are granted to different principals.
By a {\em preliminary analysis}, we can study what goals the protocol
achieves in ideal conditions where no principal acts maliciously.
An {\em empirical analysis} may follow,
whereby we can study
what goals the protocol achieves on a specific network configuration
arising from the protocol execution in the real world.
Another advantage of formalising security levels is that we can variously compare
attacks --- formally.

Our security levels belong to a finite linear order.
Protocol messages can be combined (by concatenation or encryption) or broken down (by splitting or decryption) into new messages.
We must be able to compute the security levels of the newly originated messages out of those of the message components. Therefore, we introduce a semiring whose career set is the set of security levels. Its two functions provide the necessary computational capabilities.
%
Our use of a semiring is loosely inspired to Denning's use of a lattice
to characterising secure flows of information through computer
systems~\cite{D76}.
The idea of using levels to formalise {\em access rights}
is in fact due to her.
Denning signals an attack whenever an object is
assigned a label worse than
that initially specified. We formalise protocol attacks in the same spirit.
\\

Another substantial contribution of the present work is the embedding of a novel {\em threat model} in a framework for protocol analysis. Our threat model regards {\em all principals as attackers if they
perform, {\em either deliberately or not}, any operation that is not
admitted by the protocol policy}. Crucially, it allows any number of non-colluding attackers.
This overcomes the limits of Dolev and Yao's popular threat model~\cite{dolev-yao}, which reduces a number of colluding principals to a single attacker. The example that follows shows the deeper adherence of our threat model to the real world, where anyone may attempt to subvert a protocol for his (and only his) own sake.

Let us consider Lowe's popular
attack on the asymmetric Needham-Schroeder protocol
\cite{lowefirst} within Dolev and Yao's threat model. It sees an attacker $C$
masquerade as $A$ with $B$, after $A$ initiated a session with
$C$. This scenario clearly contains an authentication attack
following the confidentiality attack whereby $C$ learns $B$'s
nonce $\Nb$ for $A$. Lowe reports that, if $B$ is a bank for
example, $C$ can steal money from $A$'s account as
follows~\cite{lowefirst}
\[
C \rightarrow B : \lbb\Na, \Nb, \mbox{``Transfer \$ 1000 from $A$'s account to $C$'s''}\rbb_\Kb
\]
where $\comp{m}_K$ stands for the ciphertext
obtained encrypting message $m$ with key $K$ (external
brackets of concatenated messages are omitted).
The bank $B$ would honour the request believing it came from
the account holder $A$.

We argue that the analysis is constrained
by the limitations of the threat model. Plunging Lowe's scenario within our
threat model highlights that $B$ has mounted an indeliberate confidentiality
attack on nonce $\Na$, which was meant to be known to $A$ and $C$ only. As $C$
did previously, $B$ can equally decide to illegally exploit his knowledge of $\Na$. If $A$ is a bank, $B$ can steal money from
$C$'s account as follows
\[
B \rightarrow A : \lbb\Na, \Nb, \mbox{``Transfer
\$ 1000 from $C$'s account to $B$'s''}\rbb_\Ka
\]
The bank $A$ would honour the request believing it came from the account
holder $C$.

The details of our findings on the Needham-Schroeder protocol can be found elsewhere~\cite{bella-bista-cambridge02}.
Our empirical analysis of the protocol uniformly detects {\em both}
attacks in terms of
decreased security levels: both $C$'s security level on $\Nb$ and
$B$'s security level on $\Na$ become lower than they would be if $C$
didn't act maliciously.
\\

The framework presented throughout this paper supersedes an existing
kernel~\cite{security-padl01,bella-bista-cambridge02} by extending it
with five substantial features.
I) The principles of the new threat model that allows all principals to behave
maliciously.
II) The combination of preliminary and empirical analyses.
III) The study of the authentication goal.
IV) The formalisation of an additional event whereby a principal discovers a secret by cryptanalysis --- this allows a larger number of network configurations to be studied through an empirical analysis.
V) A comprehensive study of how message manipulation and exposure to the network lowers the security level of the message --- this is implemented by
a new algorithm called {\sc RiskAssessment}.


Since we only deal with
bounded protocols and finite number of principals,
our framework is amenable to mechanisation by
model checking,
although this exceeds the purposes of the present paper.

\paragraph{Findings on the running example --- Kerberos.} We
demonstrate our framework on a largely deployed protocol, Kerberos.
Our preliminary analysis of the protocol formally highlights that the loss of
an {\em authorisation key} would be more serious than the loss of
a {\em service key} by showing that the former has a higher security level than
the latter. By similar means, the preliminary analysis also allows us to
compare the protocol goals in the forms they are granted to initiator and
responder. It shows that authentication of the
responder with the initiator is weaker than that of the initiator with the
responder.
To the best of our knowledge, developing such detailed observations formally
is novel to the field of protocol analysis.

The empirical analysis that follows studies an example scenario in which
a form of cryptanalysis was performed. The analysis highlights how that
event lowers a number of security levels, and so lowers confidentiality and authentication for a number of principals.

\paragraph{Paper outline.}
After an outline on semiring-based Soft Constraints Satisfaction
Problems (SCSPs)
(\S\ref{sec:soft}), our framework for protocol
analysis is described (\S\ref{framework}). Then, the
Kerberos protocol is introduced (\S\ref{sec:prot}) and analysed
(\S\ref{kerberosanalysis}).
Some conclusions (\S\ref{sec:concl}) terminate the presentation.
\section{Soft constraints}\label{sec:soft}
\label{back} Several formalisations of the concept of {\em soft
constraints} are currently available
\cite{schiex-ijcai95,fuzzy1,partial-ai,prob}. In the following, we
refer to one that is based on c-semirings
\cite{BISTATESI,jacm,toplas00}, which can be shown to generalise
and express many of the others.

A soft constraint may be seen as a constraint where each
instantiation of its variables has an associated value from a
partially ordered set. Combining constraints will then have to
take into account such additional values, and thus the formalism
has also to provide suitable operations for combination ($\times$)
and comparison (+) of tuples of values and constraints. This is
why this formalisation is based on the concept of semiring, which
is just a set plus two operations.

A semiring is a tuple $\langle A,+,\times,\0,\1 \rangle$ such
that:
\begin{itemize}
\item $A$ is a set and $\0, \1 \in A$;
\item $+$ is commutative, associative and $\0$ is its unit element;
\item $\times$ is associative, distributes over $+$, $\1$  is its
unit element and $\0$ is its absorbing element.
\end{itemize}
A {\em c-semiring} is a semiring $\langle A,+,\times,\0,\1
\rangle$ such that: $+$ is idempotent, $\1$ as its absorbing
element and $\times$ is commutative.

Let us consider the relation $\leq_S$ over $A$ such that $a \leq_S
b$ iff $a+b = b$. Then it is possible to prove that (see
\cite{jacm}):
\begin{itemize}
\item $\leq_S$ is a partial order;
\item $+$ and $\times$ are monotone on $\leq_S$;
\item $\0$ is its minimum and $\1$ its maximum;
\item $\langle A,\leq_S \rangle$ is a complete lattice and,
for all $a, b \in A$, $a+b = lub(a,b)$.
\end{itemize}
Moreover, if $\times$ is idempotent, then: $+$ distributes over
$\times$; $\langle A,\leq_S \rangle$ is a complete distributive
lattice and $\times$ its glb.

Informally, the relation $\leq_S$ gives us a way to compare (some
of the) tuples of values and constraints. In fact, when we have $a
\leq_S b$, we will say that {\em b is better than a}. Below, $a
\leq_S b$ will be often indicated by $a \leq b$.

A {\em constraint system} is a tuple $CS= \langle {\cal S},{\cal
D},{\cal V} \rangle$ where ${\cal S}$ is a c-semiring, ${\cal D}$
is a finite set (the domain of the variables) and ${\cal V}$ is an
ordered set of variables.

Given a semiring ${\cal S} = \langle A,+,\times,\0,\1 \rangle$ and
a constraint system $CS= \langle {\cal S},{\cal D},{\cal V}
\rangle$, a {\em constraint} is a pair $\langle \deff, con
\rangle$ where $con \subseteq {\cal V}$ and $\deff: {\cal
D}^{|con|} \rightarrow A$.
Therefore, a constraint specifies a set of variables (the ones in
$con$), and assigns to each tuple of values of these variables an
element of the semiring.



A soft {\em constraint problem} is a pair $\langle C, con \rangle$
where $con \subseteq {\cal V}$ and $C$ is a set of constraints:
$con$ is the set of variables of interest for the constraint set
$C$, which however may concern also variables not in $con$.

Notice that a classical CSP is a SCSP where the chosen c-semiring
is:
\[S_{CSP} = \langle \{false, true\},\vee, \wedge, false, true \rangle.\]

\begin{figure}[htbp]
\psfrag{<a> frec 0.9}{{\tiny $\langle a \rangle \rightarrow 0.9$}}
\psfrag{<b> frec 0.1}{{\tiny $\langle b \rangle \rightarrow 0.1$}}
\psfrag{<b> frec 0.5}{{\tiny $\langle b \rangle \rightarrow 0.5$}}
\psfrag{<a, a> frec 0.8}{{\tiny $\langle a,a \rangle \rightarrow
0.8$}} \psfrag{<a, b> frec 0.2}{{\tiny $\langle a,b \rangle
\rightarrow 0.2$}} \psfrag{<b, a> frec 0}{{\tiny $\langle b,a
\rangle \rightarrow 0$}} \psfrag{<b, b> frec 0}{{\tiny $\langle
b,b \rangle \rightarrow 0$}} \psfrag{c1}{{\tiny $c_1$}}
\psfrag{c2}{{\tiny $c_2$}} \psfrag{c3}{{\tiny $c_3$}} \centering
    \includegraphics[scale=.65]{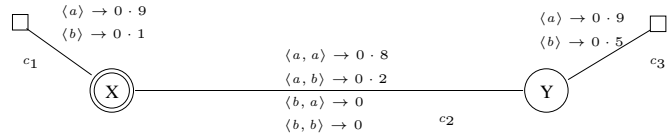}
    \caption{A fuzzy CSP}
    \label{fig:fuzzy}
\end{figure}

Fuzzy CSPs \cite{fuzzy1,fuzzy2,schiex} can instead be modelled in
the SCSP framework by choosing the c-semiring:
\[S_{FCSP} = \langle [0,1], max, min, 0, 1 \rangle.\]
\RefFig{fig:fuzzy} shows the graph representation of a fuzzy
CSP. Variables and constraints are represented respectively by
nodes and by undirected (unary for $c_1$ and $c_3$ and binary for
$c_2$) arcs, and semiring values are written to the right of the
corresponding tuples. The variables of interest (that is the set
$con$) are represented with a double circle. Here we assume that
the domain $D$ of the variables contains only elements $a$ and
$b$.

\paragraph{Combining and projecting soft constraints.} Given two
constraints $c_1 = \langle \deff_1,con_1 \rangle$ and $c_2 =
\langle \deff_2,con_2 \rangle$, their {\em combination} $c_1
\otimes c_2$ is the constraint $\langle \deff,con \rangle$ defined
by $con = con_1 \cup con_2$ and $\deff(t) = \deff_1(t
\downarrow^{con}_{con_1}) \times \deff_2(t
\downarrow^{con}_{con_2})$, where $t \downarrow^X_Y$ denotes the
tuple of values over the variables in $Y$, obtained by projecting
tuple $t$ from $X$ to $Y$.
In words, combining two constraints means building a new
constraint involving all the variables of the original ones, and
which associates to each tuple of domain values for such variables
a semiring element which is obtained by multiplying the elements
associated by the original constraints to the appropriate
subtuples.

Given a constraint $c = \langle \deff,con \rangle$ and a subset
$I$ of ${\cal V}$, the {\em projection} of $c$ over $I$, written
$c \Downarrow_I$ is the constraint $\langle \deff', con' \rangle$
where $con' = con \cap I$ and $\deff'(t') = \sum_{t / t
\downarrow^{con}_{I \cap con} = t'} \deff(t)$.
Informally, projecting means eliminating some variables. This is
done by associating to each tuple over the remaining variables a
semiring element which is the sum of the elements associated by
the original constraint to all the extensions of this tuple over
the eliminated variables.

In short, combination is performed via the multiplicative
operation of the semiring, and projection via the additive
operation.

\paragraph{Solutions.}
The {\em solution} of an SCSP problem ${\mathsf P} = \langle C,con
\rangle$ is the constraint $Sol({\mathsf P})=(\bigotimes C)
\Downarrow_{con}$.
That is, we combine all constraints, and then project over the
variables in $con$. In this way we get the constraint over $con$
which is ``induced'' by the entire SCSP.

For example, each solution of the fuzzy CSP of \RefFig{fig:fuzzy}
consists of a pair of domain values (that is, a
domain value for each of the two variables) and an associated
semiring element. Such an element is obtained by looking at the
smallest value for all the subtuples (as many as the constraints)
forming the pair. For example, for tuple $\langle a,a \rangle$
(that is, $x=y=a$), we have to compute the minimum between $0.9$
(which is the value for $x=a$), $0.8$ (which is the value for
$\langle x=a,y=a \rangle$) and $0.9$ (which is the value for
$y=a$). Hence, the resulting value for this tuple is $0.8$.

\paragraph{Partial Information and Entailment.}
A constraint is a relation
among a specified set of variables. It gives
some information on the set of possible values that those
variables may assume. Such information is usually not complete
since a constraint may be satisfied by several assignments of
values of the variables (in contrast to the situation that we have
when we consider a valuation, which tells us the only possible
assignment for a variable). Therefore, it is natural to describe
constraint systems as systems of {\em partial} information
\cite{vijay-book}.

The basic ingredients of a constraint system (defined following
the information systems idea) are a set $D*$ of {\em primitive
constraints} or {\em tokens}, each expressing some partial
information, and an entailment relation $\vdash$ defined on
$\wp(D*) \times D*$ (or its extension defined on $\wp(D*)\times
\wp(D*)$, such that $u \vdash v$ iff $u \vdash P$ for all $P \in v$) satisfying:
\begin{itemize}
\item $u \vdash P$ for all $P \in u$ (reflexivity) and
\item if $u \vdash v$ and $v \vdash z$, then $u \vdash z$ (transitivity).
\end{itemize}

As an example of entailment relation, consider $D*$ as the set of
equations over the integers; then $\vdash$ could include the pair
$\langle\{x = 3, x = y\}, y = 3\rangle$, which means that the
constraint $y = 3$ is entailed by the constraints $x = 3$ and $x =
y$. Given $X \in \wp(D*)$, let $\overline{X}$ be the set $X$ closed
under entailment. 
Then, a constraint in an information system
$\langle \wp (D*), \vdash \rangle$
is simply an element of $\overline{\wp (D)}$.

In the SCSP framework a token is simply a soft constraint (that is domain assignment
and some associated semiring values); the entailment rule will compute/change
new soft constraint (and new levels) \cite{esop2002}.

\section{Constraint Programming for Protocol Analysis}\label{framework}
This section presents our framework for
analysing security protocols. Using soft
constraints requires the definition of a c-semiring.

Our {\em security semiring} (\S\ref{secss}) is used to specify each
principal's trust on
the security of each message, that is each principal's {\em security level}
on each message.
The security levels range from the most secure (highest, greatest) level
$unknown$ to the least secure (lowest, smallest) level $public$.
Intuitively, if $A$'s security level on $m$
is $unknown$, then no principal (included $A$)
knows $m$ according to $A$, and, if $A$'s security
level on $m$ is $public$, then all principals potentially know $m$ according to $A$.
The lower $A$'s security level on $m$, the higher the number of principals
knowing $m$ according to $A$.
For simplicity, we state no relation between
the granularity of the security levels and the number of principals.

Using the security semiring, we define the {\em network constraint
system} (\S\ref{ncs}), which represents the computer network on
which the security protocols can be executed.
The development of the principals' security levels
from manipulation of the messages seen during the protocol
sessions can be formalised as a {\em security entailment}
(\S\ref{entailment}), that is an entailment relation between constraints.
Then, given a specific
protocol to analyse, we represent its assumptions
in the {\em initial SCSP}
(\S\ref{secinitialscsp}).  All admissible network configurations
arising from the protocol execution as prescribed by the protocol
designers can in turn be represented in the {\em policy SCSP}
(\S\ref{secpolicyscsp}). We also explain how to
represent any network configuration arising from the protocol
execution in
the real world as an {\em imputable SCSP} (\S\ref{impu}).

\Commento{confidentiality e authentication}
Given a security level
$l$, establishing whether our definitions of {\em
l-confidentiality} (\S\ref{conf}) or {\em l-authentication}
(\S\ref{authe}) hold in an SCSP requires calculating the
solution of the imputable SCSP and projecting it on certain principals
of interest.
The higher $l$, the stronger the goal. For example,
{\em unknown-confidentiality} is stronger than {\em
public-confidentiality}, or,
$A$'s security level on $B$'s public key (learnt via a certification authority)
being $public$ enforces
{\em public-authentication} of $B$ with $A$, which is the
weakest form of authentication.
We can also formalise confidentiality
or authentication attacks. The definitions are given within specific
methodologies of analysis.

\Commento{preliminary ed empirical}
By a {\em preliminary analysis}, we can study what goals the protocol
achieves in ideal conditions where no principal acts maliciously,
namely the very best the protocol can guarantee.
We concentrate on the policy SCSP, calculate
its solution, and project it on a principal of interest.
The process yields the principal's security levels, which allow us
to study what goals the protocol grants to that principal in ideal conditions,
and which potential attacks would be more serious than others for the
principal.
For example, the most serious confidentiality
attacks would be against those messages on which the principal
has the highest security level.

An {\em empirical analysis} may follow,
whereby we can study
what goals the protocol achieves on a specific network configuration
arising from the protocol execution in the real world. We concentrate
on the corresponding imputable SCSP,
calculate its solution and
project it on a principal of interest: we obtain the principal's
security levels on all messages.
Having done the same operations on the the policy SCSP, we can
compare the outcomes. If some level in the imputable is lower
than the corresponding level in the policy, then there is an
attack in the imputable one. In fact, some malicious activity
contributing to the network configuration modelled by the
imputable SCSP has taken place so as to lower some of the security
levels stated by the policy SCSP.

The following,
general treatment is demonstrated in~\S\ref{sec:prot}.
\subsection{The Security Semiring}\label{secss}
Let $n$ be a natural number.
We define the set $L$ of {\em security levels} as follows.
\[
L =
\{unknown,~private,~traded_1,~traded_2,~\ldots~,~traded_n,~public\}
\]
where $unknown$ is the maximum element of $L$ and $public$ is the
minimum one.

Although our security levels may appear to resemble Abadi's {\em
types}~\cite{Abadi99a}, there is in fact little similarity. Abadi
associates each message to either type $public$, or $secret$, or
$any$, whereas we define $n$ security levels with no bound on $n$,
and {\em each principal associates a level of his own to each
message} as explained in the following. Also, while Abadi's
$public$ and $private$ cannot be compared, our levels are linearly
ordered.

The security levels express each principal's trust on the security
of each message. Clearly, $unknown$ is the highest security level.
We will show how, under a given protocol, a principal assigns
$unknown$ to all messages that do not pertain to the protocol, and
to all messages that the principal does not know. A principal will
assign $private$ to all messages that, according to himself, are
known to him alone, such as his own long-term keys, the nonces
invented during the protocol execution, or any secrets discovered
by cryptanalysis. In turn, a principal will assign $traded_i$ to
the messages that are exchanged during the protocol: the higher
the index $i$, the more the messages have been handled by the
principals, and therefore the more principals have potentially
learnt those messages. So, $public$ is the lowest security level.
These security levels generalise, by the
$traded_i$ levels, the four levels that we have discussed
elsewhere~\cite{security-padl01}.

We introduce an additive operator, $+_{sec}$, and a multiplicative
operator, $\times_{sec}$. To allow for a compact
definition of the two operators, and to simplify the following
treatment, let us define a convenient double naming:
\newline
\begin{tabular}{llcl}
--&$unknown$ & $\equiv$ & $traded_{-1}$\\
--&$private$ & $\equiv$ & $traded_0$\\
--&$public$  & $\equiv$ & $traded_{n+1}$\\
\end{tabular}
\newline

Let us consider an index $i$ and an index $j$ both belonging to
the closed interval $[-1,~n+1]$ of integers. We define $+_{sec}$
and$\times_{sec}$ by the following axioms.
\begin{description}
\item[{\bf Ax. 1:}] $traded_i~+_{sec}~traded_{j}~=~traded_{min(i,j)}$
\item[{\bf Ax. 2:}] $traded_i~\times_{sec}~traded_{j}~=~traded_{max(i,j)}$
\end{description}
\begin{theorem}[Security Semiring]
The structure ${\cal S}_{sec}=\langle L, +_{sec},
\times_{sec},\,public,\, unknown\,\rangle$ is a c-semiring.
\end{theorem}
\begin{proof}[Proof hint]
Clearly, ${\cal S}_{sec}$ enjoys the same properties as the
structure $S_{finite-fuzzy}=\langle
\{-1,\ldots,n+1\},max,min,-1,n+1 \rangle$. Indeed, the security
levels can be mapped into the values in the range
${-1,\ldots,n+1}$ ($unknown$ being mapped into $0$, $public$ being
mapped into $n+1$); $+_{sec}$ can be mapped into function $max$;
$\times_{sec}$ can be mapped into function $min$. Moreover,
$S_{finite-fuzzy}$ can be proved a $c$-semiring as done with the
fuzzy semiring \cite{jacm}.
\end{proof}

Our security semiring is in fact a linear order, but the general treatment provided
here complies with the general case where $+_{sec}$ and $\times_{sec}$ must be
mapped into more complex functions than $max$ and $min$.
\subsection{The Network Constraint System}\label{ncs}
We define a constraint system $CS_n=\langle {\cal S}_{sec},{\cal
D},{\cal V} \rangle$ where:
\begin{itemize}
 \item
 ${\cal S}_{sec}$ is the security semiring (\S\ref{secss});
 \item
 ${\cal V}$ is bounded set of variables.
 \item
 ${\cal D}$ is an bounded set of values including
 the empty message $\comp{}$ and all atomic messages,
 as well as all messages recursively obtained by concatenation
 and encryption.

\end{itemize}
We name $CS_n$ as {\em network constraint system}.
The elements of $\cal V$ stand for the network principals, and the
elements of ${\cal D}$ represent all possible messages.
Atomic messages typically are principal names, timestamps, nonces and
cryptographic keys. Concatenation and encryption operations can be
applied a bounded number of times.

Notice that $CS_n$ does not depend on any protocols, for it merely
portrays a computer network on which any protocol
can be implemented. Members of ${\cal V}$ will be indicated by
capital letters, while members of ${\cal D}$ will be in small
letters.
\subsection{Computing the Security Levels by Entailment}\label{entailment}
Recall that each principal associates his own security levels to the
messages. Those levels evolve while the principal
participates in the protocol and performs off-line operations such as
encryption, concatenation, decryption, and splitting.
We define four rules to compute
the security levels that each principal gives to the newly generated messages.
The rules are presented in \RefFig{fig:rules}, where function $\deff$ is
associated to a generic constraint projected on a generic
principal $A$.

\begin{figure}[ht]
\small{
\begin{tabbing}
{\bf Encryption:}\\ \hspace{2mm}
$\irule{
\deff(m_1)=v_1;~~\deff(m_2)=v_2;~~\deff(\comp{m_1}_{m_2})=v_3}
{\deff(\comp{m_1}_{m_2})=(v_1+_{sec}v_2)\times_{sec}v_3}$\\
\end{tabbing}
\begin{tabbing}
{\bf Concatenation:}\\ \hspace{3mm}
$\irule{
\deff(m_1)=v_1;~~\deff(m_2)=v_2;~~\deff(\comp{m_1,m_2})=v_3;}
{\deff(\comp{m_1,m_2})=(v_1+_{sec}v_2)\times_{sec}v_3}$\\
\end{tabbing}
\begin{tabbing}
{\bf Decryption:}\\ \hspace{3mm}
$\irule{
\deff(m_1)=v_1;~~\deff(m_2^{-1})=v_2;~~\deff(\comp{m_1}_{m_2})=v_3;~~v_2,v_3<unknown}
{\deff(m_1)=v_1\times_{sec}v_2\times_{sec}v_3}$\\
\end{tabbing}
\begin{tabbing}
{\bf Splitting:}\\ \hspace{3mm}
$\irule{
\deff(m_1)=v_1;~~\deff(m_2)=v_2;~~\deff(\comp{m_1,m_2})=v_3}
{\deff(m_1)=v_1\times_{sec}v_3;~~\deff(m_2)=v_2\times_{sec}v_3}$
\end{tabbing}
}
\caption{Computation rules for security levels.}
\label{fig:rules}
\end{figure}

A message is known to a principal when the principal's security level on
that message is lower than $unknown$.
Encryption and concatenation build up new messages from known ones.
The new messages must not get a worse security level than the known ones
have. So, the corresponding rules choose the better of the given levels.
Precisely, if messages $m_1$ and $m_2$
have security levels $v_1$ and $v_2$ respectively, then the
encrypted message $\comp{m_1}_{m_2}$ and the compound message
$\comp{m_1, m_2}$, whose current level be some $v_3$,
get a new level that is the better of $v_1$
and $v_2$, ``normalised'' by $v_3$.
This normalisation, which is done in terms of the
$\times_{sec}$ operator, influences the result only if the new level
is better than the current level.

Decryption and splitting break
down known messages into new ones.
The new messages must not get a better security level than the known ones
have.
So, the corresponding rules choose the worse of the given levels
by suitable applications of
$\times_{sec}$, and assign it to the new messages. Recall that,
in case of asymmetric cryptography, the decryption key for a
ciphertext is the inverse of the key that was used to create the
ciphertext. So the rule for decryption considers the inverse of
message $m_2$ and indicates it as $m_2^{-1}$. Conversely, in case
of symmetric cryptography, we have $m_2^{-1} = m_2$. The rule for
splitting presupposes that concatenation is transparent in the
sense that, for any index $n$, an $n$-component message can be
seen as a 2-component message, namely
$
\comp{m_1, m_2, \ldots, m_n} = \comp{m_1, \comp{m_2, \ldots, m_n}}
$.
We now define a binary relation between constraints.

\begin{definition}[Relation $\ent$]
\label{def:ent} Consider two constraints $c_1,c_2\in C$ such that
$c_1=\langle \deff_1, con\rangle$ and $c_2=\langle \deff_2,
con\rangle$. The binary relation $\ent$ is such that
$c_1 \ent c_2$ iff $\deff_2$ can be obtained from $\deff_1$ by a number
(possibly zero) of applications of
the rules in \RefFig{fig:rules} .
\end{definition}

\begin{theorem}[Relation $\ent$ as entailment relation]\label{thm:ent}
The binary relation $\ent$ is an entailment relation.
\end{theorem}
\begin{proof}[Proof hint.]
Relation $\ent$ enjoys the reflexivity and
transitivity properties that are needed to be an entailment
relation.
\end{proof}

In the following, $c^{\ent}$ represents the reflexive, transitive closure of
the entailment relation $\ent$ applied to the constraint $c$.
While other entailment relations (e.g.~\cite{esop2002})
involve all constraints that are related by the partial order $\leq_S$,
the security entailment only concerns the subset of those constraints obtainable
by application of the four rules in \RefFig{fig:rules}.

\subsection{The Initial SCSP}\label{secinitialscsp}
The designer of a protocol must also develop
a {\em policy} to accompany the protocol.
The policy for a protocol $\cal P$ is a set of rules
stating, among other things,
the preconditions necessary for the protocol execution,
such as which messages are public, and which messages are private for
which principals.

It is intuitive to capture these policy rules by our security
levels (\S\ref{secss}).
Precisely, these rules can be translated into unary constraints.
For each principal $A \in {\cal V}$, we define a unary constraint that
states $A$'s security levels as follows. It
associates security level $public$ to those messages that
are known to all, typically
principal names and timestamps; level $private$ to $A$'s initial
secrets, such as keys
(e.g., $A$'s long-term key if $\cal P$ uses symmetric
cryptography, or $A$'s private key if $\cal P$ uses asymmetric
cryptography, or $A$'s pin if $\cal P$ uses smart cards) or
nonces;
level $unknown$ to all remaining domain values (including, e.g.,
the secrets that $A$ will invent during the protocol execution,
or other principals' initial secrets).

This procedure defines what we name {\em initial SCSP for} $\cal
P$, which specifies the principals' security levels when no
session of $\cal P$ has yet started. Notice that the constraint
store representing each principal's security levels is computed
using the reflexive, transitive, closure of the entailment
relation (\S\ref{entailment}). So, when a new message is invented,
the corresponding constraint
is added to the store along with all constraints that can be
extracted by entailment.

Considerations on how official protocol specifications often fail
to provide a satisfactory policy~\cite{miojsac} exceed the scope of
this paper. Nevertheless, having to define the initial SCSP for a
protocol may help pinpoint unknown deficiencies or ambiguities in
the policy.

\subsection{The Policy SCSP}\label{secpolicyscsp}
The policy for a
protocol $\cal P$ also establishes which messages must be
exchanged during a session between a pair of principals while
no-one performs malicious activity. The protocol designer
typically writes a single step as $A \rightarrow B : m$, meaning
that principal $A$ sends message $m$ to principal $B$. The policy
typically allows each principal to participate in a number of
protocol sessions inventing a number of fresh messages.
Assuming both these numbers to be bounded, a
bounded number of {\em events} may take
place~\cite{nancymitchell}.
Because no principal is assumed to be acting maliciously, no message is
intercepted, so a message that is sent is certain to reach its intended
recipient. Therefore, we only formalise the two following events.
\begin{enumerate}
\item A principal invents a fresh message (typically a new nonce).
\item A principal sends a message (constructed by some sequence of applications
of encryption, concatenation, decryption, and splitting) to another principal,
and the message is delivered correctly.
\end{enumerate}

Clearly, additional events can be formalised to capture protocol-specific
details, such as principal's annotation of sensitive messages,
message broadcast, SSL-secure trasmission, and so on.

We read
from the protocol policy each allowed step of the form $A \rightarrow B :
m$ and its informal description, which explains whether $A$
invents $m$ or part of it. Then, we build the {\em policy SCSP
for} $\cal P$ by the algorithm in \RefFig{procedure}.

\begin{figure}[h]
{\footnotesize
\hspace*{-78mm}
{\sc BuildPolicySCSP($\cal P$)}
\vspace*{-1mm}
\begin{center}
\begin{tabular}{ll}
  1.&\quad {\sf p} $\leftarrow$ {\em initial SCSP for} $\cal P$;\\
  2.&\quad {\bf for} each event $ev$ allowed by the policy for $\cal P$ {\bf do}\\
  3.&\quad\quad {\bf if} $ev~=~(A$ {\em invents} $n$, for some $A$
and $n$) {\bf then}\\
  4.&\quad\quad\quad
           {\sf p} $\leftarrow$ {\sf p} extended with unary constraint
on $A$
              that assigns\\
        &\quad\quad\quad\quad\hspace*{3mm}$private$ to $n$ and $unknown$ to all
other messages;\\
 5.&\quad\quad {\bf if} $ev~=~(A$ {\em sends} $m$ {\em to} $B$ {\em
not intercepted}, for some $A$, $m$ and $B$) {\bf then} \\
 6.&\quad\quad\quad $c \leftarrow Sol({\mathsf
p})\Downarrow_{\{A\}}$; \\
 7.&\quad\quad\quad {\bf let} $\langle \deff, con\rangle = c^{\ent}$ {\bf in}
 $newlevel \leftarrow$ {\sc RiskAssessment($\deff(m)$)};\\
   8.&\quad\quad\quad
           {\sf p} $\leftarrow$ {\sf p} extended with binary constraint
              between $A$ and $B$ that assigns\\
        &\quad\quad\quad\quad\quad$newlevel$ to
$\langle \comp{}, m \rangle$ and $unknown$ to all other tuples;\\
  9.&\quad return {\sf p};
\end{tabular}
\end{center}
} \caption{Algorithm to construct the policy SCSP for a protocol $\cal
P$.}\label{procedure}
\end{figure}
The algorithm considers the initial SCSP (line 1) and extends it with
new constraints induced by each of the events occurring during the
protocol execution (line 2).
If the current event is a principal $A$'s
inventing a message $n$ (line 3), then a unary constraint is added on
variable $A$ assigning security level $private$ to the domain
value $n$, and $unknown$ to all other values (line 4). If that event is a
principal $A$'s sending a message $m$ to a principal $B$ (line 5), then
the solution of the current SCSP ${\mathsf p}$ is computed and
projected on the sender variable $A$ (line 6),
and extended by entailment (line 7). The last two steps yield
$A$'s view of the
network traffic. In particular, also $A$'s security level on $m$ is updated
by entailment. For example, if $m$ is built as $\comp{\Na, \Nb}$,
the security levels of $\Na$ and $\Nb$ derive from the computed solution,
and then the level of $m$ is obtained by the concatenation rule of
the entailment relation.

At this stage, $A$'s security level on $m$ is updated again by
algorithm {\sc RiskAssessment} (line 7).
As explained in the next section,
this shall assess the risks that $m$ runs following $A$'s manipulation and the exposure to the network.
The current SCSP can be now extended with
a binary constraint on the pair of variables $A$ and $B$ (line 8). It assigns
the newly computed security level $newlevel$ to the tuple $\langle \comp{},
m \rangle$ and $unknown$ to all other tuples.
This reasoning is repeated for each of the bounded number of
events allowed by the policy.
When there are no more events to process,
the current SCSP is returned as policy SCSP for $\cal P$ (step 9), which is our
formal model for the idealised protocol.
Termination of the algorithm is guaranteed by finiteness of the
number of allowed events. Its complexity is clearly linear in the
number of allowed events, which is in turn exponential in the
length of the exchanged messages~\cite{nancymitchell}.

We remark that a binary constraint between (a pair of variables formalising respectively)
sender and receiver of $m$, which assigns some newly computed security level to the tuple
$\langle \comp{}, m \rangle$  confirms that the receiver's level on $m$ is influenced by
the event that the constraint formalises, as opposed to the sender's level which is not.

\subsection{Assessing the Expected Risk}
Each network event involves some message.
The events that expose their messages to the network, such as
to send or receive or broadcast a message, clearly impose some {\em expected
risk} on those messages --- ideal message security is never to use that
message.
The {\em risk function} $\rho$ expresses how the expected
risk affects the security levels of the messages that are involved.

The actual
definition of the risk function depends on the protocol policy, which should
clearly state the expected risk for each network event when the protocol
is executed in its intended environment. But ``often protocols are used in
environments other than the ones for which they were originally
intended''~\cite{meaprivate}, so the definition
also depends on the specific environment that is considered.

The risk function should take as parameters
the given security level and the network event that is influencing that level.
The second parameter can be omitted for simplicity from this presentation
because of the limited number of events we are modelling.
Indeed,
we will only have to
compute the function
for the network event whereby
a message is sent on the network (either intercepted or not), whereas
if we modelled, for example, a broadcast
event, then the assessment for that particular
event would have to yield $public$.

The risk function must enjoy the two following properties.
\begin{enumerate}
\item[i.] {\em Extensivity}.
This property means that $\rho(l) \leq l$ for any $l$. It captures the
requirement that each manipulation of a message decrease its security level
--- each manipulation increases the risk of tampering.
\item[ii.] {\em Monotonicity}.
This property means that $l_1 \leq l_2$ implies $\rho(l_1) \leq \rho(l_2)$ for any $l_1$ and $l_2$. It captures the requirement that the expected risk preserve the $\leq$ relation between
any pair of given security levels.
\end{enumerate}

Notice that
we have stated no
restrictions on the values of the risk function. Therefore,
an initial total order, e.g. $l_1 < l_2$, may at times be preserved,
such as $\rho(l_1) < \rho(l_2)$, or
at other times be
hidden, such as $\rho(l_1) = \rho(l_2)$.

As a simple example of risk function we choose the
following variant of the predecessor function.
It takes a security level and produces its predecessor in the linear order
induced by $+_{sec}$ on the set $L$ of security levels, unless the given level
is the lowest, $public$, in which case the function leaves it unchanged.
Our algorithm {\sc RiskAssessment}
in general serves to implement the risk function.
Figure~\ref{fig:minusone} shows the algorithm for our example function.

\begin{figure}[h]
{\footnotesize
\hspace*{-32mm}
{\sc RiskAssessment($l$)}
\vspace*{-1mm}
\begin{center}
\begin{tabular}{ll}
1.&\quad {\bf let} $traded_i = l$ {\bf in}\\
2.&\quad\quad {\bf if} $i=n+1$ {\bf then} $l'
\leftarrow l$\\
3.&\quad\quad\quad\quad\quad\quad\quad\hspace*{2mm}{\bf else}~~$l' \leftarrow traded_{i+1}$;\\
4.&\quad return $l'$;
\end{tabular}
\end{center}
} \caption{Implementation for a simple risk function.}
\label{fig:minusone}
\end{figure}

We remark that all considerations we advance in the sequel of this paper
merely rely on the two properties we have required for a risk function and
are therefore independent from the specific example function.
However, the protocol
analyser may take, depending on his focus,
more detailed risk functions,
such as for checking originator(s) or recipient(s) of the current event
(conventional principals, trusted third principals, proxi principals, etc.),
the network where it is being performed (wired or wireless), and so on.

One could think of embedding the risk function at the constraint level rather
than at a meta-level as we have done. That would be
possible by embedding the appropriate refinements in the entailment rules.
For example, let us consider an agent's
construction of a message $m=\comp{m_1,m_2}$, which is currently $traded_{i}$,
from concatenation of $m_1$ and $m_2$, which are
$traded_{i_1}$ and $traded_{i_2}$ respectively.
The entailment rule should first compute the
maximum between the levels of the components
(that is the minimum between the indexes),
obtaining $traded_{min(i_1,i_2)}$.
Then, it should compute the minimum between the level just computed and that
of $m$ (that is the maximum between the indexes),
obtaining $traded_{max(min(i_1,i_2),i)}$.
Finally, the rule should apply the risk function. With our example risk
function, it should yield $traded_{max(min(i_1,i_2),i)+1}$.
But the security levels would be decremented every time the
entailment relation were applied. This would violate a general requirement
of constraint programming, that is $c^{\ent}={c^{\ent}}^{\ent}$.
Hence, the decrement at the meta level is preferable.

\subsection{The Imputable SCSPs}\label{impu}
A real-world network history induced by a protocol $\cal P$ must
account for malicious activity by some principals.
Each such history can be viewed as a sequence of events of four
different forms.
\begin{enumerate}
\item A principal invents a fresh message (typically a new nonce).
\item A principal sends a message (constructed by some sequence of applications
of encryption, concatenation, decryption, and splitting) to another principal,
and the message is delivered correctly.
\item A principal sends a message (constructed as in the previous event) to
another principal, but a third principal intercepts it.
\item A principal discovers a message by cryptanalysing another message.
\end{enumerate}

Unlike the first two events, which were formalised also for constructing
the policy SCSP, the last two are new, as they are outcome of malicious
activity. We remark that the third event signifies that the message reaches
some unexpected principal rather than its intended recipient.

We can model any network configuration at a certain point in any
real-world network history as an SCSP by modifying the
algorithm given in \RefFig{procedure}
as in \RefFig{procextens} (unmodified fragments are omitted).
The new algorithm takes as inputs a protocol $\cal P$
and a network configuration $nc$
originated from the protocol execution.
The third type of event is processed as follows:
when a message is sent by $A$ to $B$ and is
intercepted by another principal $C$,
the corresponding constraint must be stated on the
pair $A, C$ rather than $A, B$.
The fourth type of event is processed by stating a unary constraint
that assigns $private$ to the
cryptanalyser's security level on the discovered message.

\begin{figure}[h]
{\footnotesize
\hspace*{-72mm}
{\sc BuildImputableSCSP($\cal P$, $nc$)}
\vspace*{-4mm}
\begin{center}
\begin{tabular}{ll}
    &\vdots\\[.5ex]
  2.&{\bf for} each event $ev$ in $nc$ {\bf do}\\
    &\quad\vdots\\[.5ex]
 8.1.&\quad{\bf if} $ev~=~(A$ {\em sends} $m$ {\em to} $B$
{\em intercepted by} $C$, for some $A$, $m, B$ and $C$)  {\bf then} \\
 8.2.&\quad\quad$c \leftarrow Sol({\mathsf
p})\Downarrow_{\{A\}}$; \\
 8.3.&\quad\quad {\bf let} $\langle \deff, con\rangle = c^{\ent}$ {\bf in}
 $newlevel \leftarrow$ {\sc RiskAssessment($\deff(m)$)};\\
 8.4.&\quad\quad
           {\sf p} $\leftarrow$ {\sf p} extended with binary constraint
              betweeen $A$ and $C$ that assigns\\
        &\quad\quad\quad\hspace*{3mm}$newlevel$ to
$\langle \comp{}, m \rangle$ and $unknown$ to all other tuples;\\
 8.5.&\quad{\bf if} $ev~=~(C$ {\em cryptanalyses} $n$ {\em from} $m$,
     for some $C$, $m$ and $n$) {\bf then}\\
 8.6.&\quad\quad
           {\sf p} $\leftarrow$ {\sf p} extended with unary constraint
              on $C$ that assigns\\
        &\quad\quad\quad\hspace*{3mm}$private$ to
  $n$ and $unknown$ to all other messages;\\
    &\quad\vdots
\end{tabular}
\end{center}
} \caption{Algorithm to construct an imputable SCSP for $\cal P$
(fragment).}\label{procextens}
\end{figure}

The new algorithm outputs what we name an {\em imputable SCSP} for
$\cal P$. Both the initial SCSP and the policy SCSP may be
viewed as imputable SCSPs.
Because we have assumed all our objects to be bounded,
the number of possible network configurations
is bounded and so is the number of imputable
SCSPs for $\cal P$.
\subsection{Formalising Confidentiality}\label{conf}
``{\em Confidentiality is the protection of information from
disclosure to those not intended to receive it}''~\cite{neuman96kerberos}.
This definition is often simplified into one that is easier to
formalise within Dolev-Yao's~\cite{dolev-yao} model with a single attacker:
a message is confidential if it is not known to the
attacker. The latter definition is somewhat
weaker: if a principal $C$ who is not the attacker gets to
learn a session key for $A$ and $B$, the latter definition holds
but the former does not.
To capture the former definition, we adopt the following threat model:
{\em all principals are attackers if they
perform, {\em either deliberately or not}, any operation that is not
admitted by the protocol policy}. As we have discussed in the introduction
to this paper, our threat model exceeds the limits of Dolev-Yao's by allowing
us to analyse scenarios with an unspecified number of non-colluding attackers.

A formal definition of confidentiality should account for the
variety of requirements that can be stated by the protocol policy.
For example, a message might be required to remain confidential
during the early stages of a protocol but its loss during the late
stages might be tolerated, as is the case with SET~\cite{miojsac}.
That protocol typically uses a fresh session key to transfer some
certificate once, so the key loses its importance after the
transfer terminates.

Another possible requirement is that certain messages, such as those
signed by a {\em root certification authority} to associate
the principals to their public keys~\cite{miojsac}, be entirely
reliable. Hence, at least those messages must be assumed to
be safe from cryptanalysis.
%
Also, a protocol may give different guarantees about
its goals to different principals~\cite{miophd},
so our definition of confidentiality must depend on the
specific principal that is considered.

Using the security levels, we
develop uniform definitions of confidentiality and of confidentiality
attack that account for any policy requirement.
Intuitively, if
a principal's security level on a message is $l$, then the message is
{\em l-confidential} for the principal because the security level
in fact formalises
the principal's trust on the security, meant as confidentiality, of the
message (see the beginning of \S\ref{framework}).
Thus, if an imputable SCSP features a principal with a lower security
level on a message w.r.t. the corresponding level in the policy SCSP,
then that imputable SCSP bears a {\em confidentiality attack}.

Here, $l$ denotes a generic security level, $m$ a generic
message, $A$ a generic principal.
Also, {\sf P} indicates the policy SCSP for a
generic security protocol, and {\sf p} and {\sf p}$'$ some
imputable SCSPs for the same protocol.
We define $Sol({\mathsf P})\Downarrow_{\{A\}} = \langle
\Deff_A, \{A\} \rangle$,
$Sol({\mathsf p})\Downarrow_{\{A\}}
= \langle \deff_A, \{A\}\rangle$, and $Sol({\mathsf
p}\,')\Downarrow_{\{A\}} = \langle \deff\,'_A, \{A\}\rangle$.

\begin{definition}[$l$-confidentiality]\label{deflconf}
$l$-confidentiality of $m$ for $A$ in $\mathsf p$
$\Longleftrightarrow$ $\deff_A(m) = l$.
\end{definition}

\subsubsection{Preliminary analysis of confidentiality}
\label{preliminaryconf}
The preliminary analysis of the confidentiality goal
can be conducted on the policy SCSP for the given protocol.

Let us calculate the solution of the policy SCSP, and project it on some
principal $A$. Let us suppose that two messages $m$ and $m'$ get
security levels $l$ and $l'$ respectively, $l' < l$.
Thus, even if no principal acts maliciously,
$m'$ must be manipulated more than $m$, so
$A$ trusts that $m'$ will be more at risk than $m$.
We can conclude that the protocol achieves a {\em stronger confidentiality
goal on $m$ than on $m'$} even if it is executed in ideal conditions.
Also, $m$ may be used
to encrypt $m'$, as is the case with Kerberos (\S\ref{kerconf}) for example.
Therefore,
losing $m$ to a malicious principal would be more serious
than losing $m'$. We address a principal's
loss of $m$ as {\em confidentiality
attack on m}. A more formal definition of confidentiality attack
cannot be given within the preliminary analysis because no malicious activity
is formalised. So, the following definition concerns potential
confidentiality attacks that may occur during the execution

\begin{definition}[Potential, worse confidentiality attack]\label{pre-conf-att}
Suppose that there is $l$-confidentiality of $m$ in {\sf P} for $A$, that
there is $l'$-confidentiality of $m'$ in {\sf P} for $A$, and that $l' < l$;
then, {\em a confidentiality
attack on $m$ would be worse than a confidentiality attack on $m'$}.
\end{definition}

\subsubsection{Empirical analysis of confidentiality}
\label{empiricalconf}
By an empirical analysis, we consider a specific real-world scenario arising
from the execution of a protocol and build the corresponding imputable
SCSP {\sf p}.
If the imputable SCSP achieves a weaker confidentiality goal of some message
for some principal than the policy SCSP does,
then the principal has mounted, either deliberately or not, a confidentiality
attack on the message.
\begin{definition}[Confidentiality
attack]\label{emp-conf-att}
Confidentiality attack by $A$ on $m$ in
{\sf p} $\Longleftrightarrow$ $l$-confidentiality of $m$ in {\sf
P} for $A$~~$\wedge$~~$l'$-confidentiality of $m$ in {\sf p} for
$A$~~$\wedge$~~$l' < l$.
\end{definition}

Therefore, there is a confidentiality attack by $A$ on $m$ in {\sf p}
iff $\deff_A (m) < \Deff_A (m)$.
The more an attack lowers a security level, the worse that attack, so
confidentiality attacks can be variously compared.
%
%
%
For example, let us
consider two confidentiality attacks by some agent on a message.
If the message is $l$-confidential for the agent in the policy SCSP, but
is $l'$-confidential and $l''$-confidential respectively in some imputable
SCSPs {\sf p} and {\sf p}$'$ for the same agent, then
$l > l' > l''$ implies that the
attack mounted in {\sf p}$'$ is worse than that in {\sf p}.
Likewise, let us consider
two messages $m$ and $m'$ that are both $l$-confidential for some
agent in the policy SCSP. If $m$ is $l'$-confidential, and $m'$ is
$l''$-confidential in {\sf p}, then $l > l' > l''$ implies that the
attack mounted on $m'$ is worse than that on $m$.

\subsection{Formalising Authentication}\label{authe}
The authentication goal enforces the principals' presence in the network
and possibly their participation in specific protocol sessions.
It is achieved by means of messages that ``{\em speak
about}'' principals. For example, in a symmetric cryptography setting,
given a session key $\Kab$ relative to the session between principals
$A$ and $B$ and known to both, message $\comp{A,\Na}_\Kab$
received by $B$ informs him that $A$ is running the session based
on nonce $\Na$ and key $\Kab$, namely the message authenticates $A$ with $B$.
An equivalent message in an
asymmetric setting could be $\comp{\Nb}_{\Ka^{-1}}$, which $B$ can decrypt
using $A$'s public key.
Also $B$'s mere knowledge of $\Ka$ as being
$A$'s public key is a form of authentication of $A$ with $B$. Indeed, $A$
must be a legitimate principal because $\Ka$
is typically certified by a certificate of the form
$\comp{A, \Ka}_{\mathit K_{ca}}$, ${\mathit K_{ca}}$ being the
public key of a certification authority.
It follows that
security protocols may use a large variety of message forms to achieve the
authentication goal --- the ISO standard in fact
does not state a single form to use~\cite{isoosi}.

In consequence, we declare a predicate $speaksabout(m,A)$, but
do not provide a formal definition for it because this would
necessarily have to be restrictive. However, the examples above
provide the intuition of its semantics. There is {\em
l-authentication} of $B$ with $A$ if
there exists a message such that $A$'s security level on it is $l$,
and the message speaks about $B$.
This signifies that $A$ received a message conveying $B$'s aliveness.

\Commento{ ho inserito per avercele in commento le 3 definizioni
\\
 l-aliveness of $B$ with $A$ via $m$ in {\sf p}
$\Longleftrightarrow~\exists m$ s.t. $def_A(m) = l < unknown$ and
$speaksabout(m,B),~\exists m'$ s.t. $\deff_B(m') < unknown$.
\\
 l-weakagreement of $B$ with $A$ via $m$ in {\sf p}
$\Longleftrightarrow~\exists m$ s.t. $def_A(m) = l < unknown$ and
$speaksabout(m,B),~\exists m'$ s.t. $speaksabout(m',B)$ and
$\deff_B(m') < unknown$.
\\
}

\begin{definition}[l-authentication]\label{deflauth}
l-authentication of $B$ with $A$ in {\sf p}
$\Longleftrightarrow~\exists~m$ s.t. $def_A(m) = l < unknown
\wedge speaksabout(m,B)~\wedge~\deff_B(m) < unknown$.
\end{definition}

The definition says that there is $l$-authentication of $B$ with $A$ whenever
both $A$ and $B$'s security levels on a message that speaks about $B$ are
less than unknown, $l$ being $A$'s level on the message.
The intuition behind the definition is that messages that $B$ sends $A$ for
authentication will produce a strong level of authentication if they reach
$A$ without anyone else's tampering. Otherwise the level of authentication
gets weaker and weaker. Precisely,
the lower $A$'s security level on $m$, the weaker the
authentication of $B$ with $A$.

Weaker forms of authentication hold when, for example, $B$ sends a
message speaking about himself via a trusted third principal, or when a
malicious principal overhears the message (recall that each event of sending
decreases the security level of the sent message).
Our definition applies uniformly to both circumstances by the appropriate
security level.

Another observation is that the weakest form, $public$-authentication, holds
for example of $B$ with $A$ in an asymmetric-cryptography
setting by the certificate for $B$'s public key in any imputable SCSP where $A$
received the certificate.
Likewise, the spy could always forge a
$public$ message that speaks about $B$, e.g. a message
containing $B$'s identity. But in fact $public$-authentication always holds
between any pairs of principals because principals' names are known to all.

\subsubsection{Preliminary analysis of authentication}
\label{preliminaryauth}
As done with the confidentiality goal (\S\ref{preliminaryconf}), the
preliminary
analysis of the authentication goal can be conducted on the policy
SCSP for the given protocol.

Once we calculate the solution of that SCSP, we can apply our definition
of $l$-authentication, and verify what form
of authentication is achieved. In particular, if there is $l$-authentication
of $B$ with $A$, and $l'$-authentication of $D$ with $C$, $l' < l$,
then we can conclude that the protocol achieves a {\em stronger authentication
goal of $B$ with $A$, than of $D$ with $C$}.
We address a principal's masquerading as $B$ with $A$
as {\em authentication attack on} $A$ {\em by means of} $B$.
A more formal definition of authentication attack cannot be given at this
stage, since no principal acts maliciously in the policy SCSP,
However, we can compare potential authentication attacks in case
they happen during the protocol execution.

\begin{definition}[Potential, worse authentication attack]\label{pre-auth-att}
Suppose that there is $l$-authentication of $B$ with $A$ by $m$ in {\sf P},
that there is $l'$-authentication of $D$ with $C$ by $m'$ in {\sf P},
and that $l' < l$; then {\em an authentication attack on $A$ by means of $B$
would be worse than an authentication attack on $C$ by means of $D$}.
\end{definition}

\subsubsection{Empirical analysis of authentication}
\label{empiricalauth}
If the policy SCSP {\sf P} achieves $l$-authentication of $B$ with $A$ by
$m$, and an imputable SCSP {\sf p}
achieves a weaker form of authentication
between the same principals by the same message, then the latter
SCSP bears an authentication attack.

\begin{definition}[Authentication attack]\label{emp-auth-att}
Authentication attack on $A$ by means of $B$ in {\sf p}
$\Longleftrightarrow$ $l$-authentication of $B$ with $A$ in {\sf
P} $\wedge$ $l'$-authentication of $B$ with $A$ in {\sf p}
$\wedge~~l' < l$.
\end{definition}

If a malicious principal has intercepted a message $m$
that authenticates $B$ with $A$, and forwarded $m$ to $B$ in some imputable
SCSP {\sf p}, then, according to the previous definition,
there is an authentication attack on $A$ by means of $B$ in {\sf p}.

\section{The Kerberos Protocol}\label{sec:prot}
Kerberos is a protocol based on symmetric cryptography meant to
distribute session keys with authentication over local area
networks. The protocol has been developed in several variants
(e.g.~\cite{techplan}), and also integrated with smart
cards~\cite{kerberossmart}. Here, we refer to the version by Bella
and Riccobene~\cite{miojucs}.

\begin{figure}[htbp]
\centering
    \includegraphics[scale=.7]{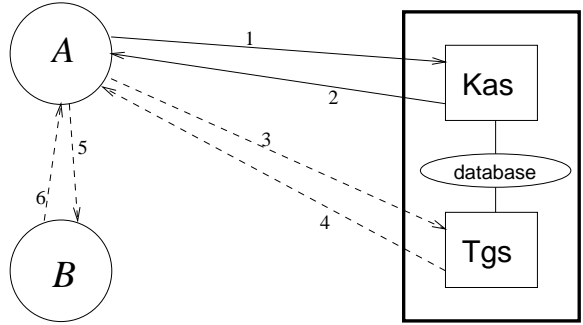}
    \caption{The Kerberos layout.}
    \label{layout}
\end{figure}

The layout in \RefFig{layout} shows that Kerberos relies on
two servers, the {\em Kerberos Authentication Server} ($\kas$ in
brief), and the {\em Ticket Granting Server} ($\tgs$ in brief).
The two servers are trusted, namely they are assumed to be secure
from the spy's tampering. They have access to an internal database
containing the long-term keys of all principals. The database
is in turn assumed to be secure. Only the first two steps of the
protocol are mandatory, corresponding to a principal $A$'s
authentication with $\kas$. The remaining steps are optional as
they are executed only when $A$ requires access to a network
resource $B$.
\begin{figure}[ht]
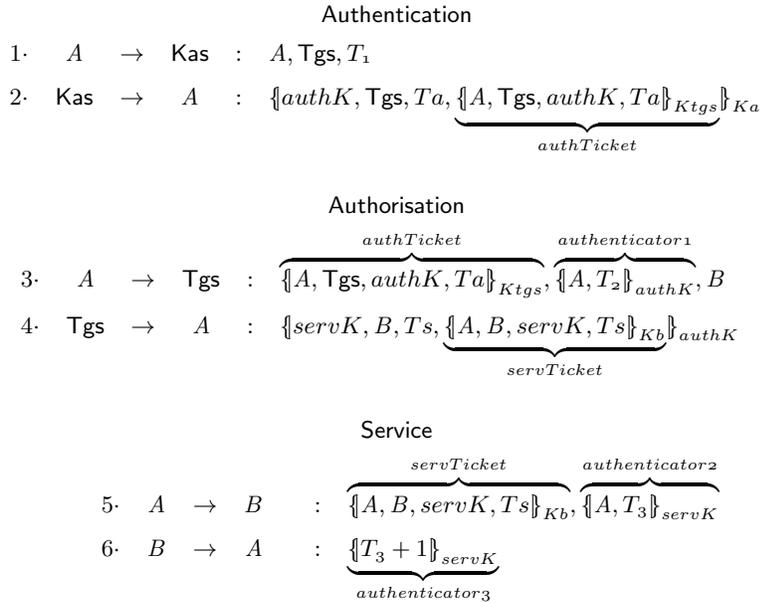

{\sf Authentication}\\[1.5ex]
$
\begin{array}{llcccll}
  &1. & A & \to & \kas & : & A,\tgs,\Tuno \vspace{2mm}\\
  &2. & \kas & \to & A & : &\comp{\ak,\tgs,\Ta,\underbrace{\comp{A,\tgs,\ak,\Ta}_{\Ktgs}}_{\at}}_{\Ka}
\end{array}
$
\newline
\newline
\newline
{\sf Authorisation}\\[1.5ex]
$
\begin{array}{llcccll}
  &3. & A & \to & \tgs & : &\overbrace{\comp{A,\tgs,\ak,\Ta}_{\Ktgs}}^{\at}, \overbrace{\comp{A,\Tdue}_{\ak}}^{\authenticatoruno},B \vspace{2mm}\\
  &4. & \tgs & \to & A & : &\comp{\sk,B,\Tt,\underbrace{\comp{A,B,\sk,\Tt}_{\Kb}}_{\st}}_{\ak}
\end{array}
$
\newline
\newline
\newline
{\sf Service}\\[1.5ex]
$
\begin{array}{llcccll}
  &5. & A & \to & B &\quad : &\overbrace{\comp{A,B,\sk,\Tt}_{\Kb}}^{\st}, \overbrace{\comp{A,\Ttre}_{\sk}}^{\authenticatordue} \vspace{2mm}\\
  &6. & B & \to & A &\quad : &\underbrace{\comp{\Ttre +1}_{\sk}}_{\authenticatortre}
\end{array}
$
\caption{The Kerberos protocol.} \label{fig:kerberos_prot}
\end{figure}

In the {\sf authentication} phase, the initiator $A$ queries
$\kas$ with her identity, $\tgs$ and a timestamp $\Tuno$; $\kas$
invents a session key and looks up $A$'s shared key in the
database. It replies with a message sealed by $A$'s shared key
containing the session key, its timestamp $\Ta$, $\tgs$ and a
ticket. The session key and the ticket are the credentials to use
in the subsequent authorisation phase, so we address them as {\em
authkey} and {\em authticket} respectively.

Now, $A$ may start the {\sf authorisation} phase. She sends $\tgs$
a three-component message including the authticket, an
authenticator sealed by the authkey containing her identity and a
new timestamp $\Tdue$, and $B$'s identity. The lifetime of an
authenticator is a few minutes. Upon reception of the message,
$\tgs$ decrypts the authticket, extracts the authkey and checks
the validity of its timestamp $\Ta$, namely that $\Ta$ is not too
old with respect to the lifetime of authkeys. Then, $\tgs$
decrypts the authenticator using the authkey and checks the
validity of $\Tdue$ with respect to the lifetime of
authenticators. Finally, $\tgs$ invents a new session key and
looks up $B$'s shared key in the database. It replies with a
message sealed by the authkey containing the new session key, its
timestamp $\Ts$, $B$ and a ticket. The session key and the ticket
are the credentials to use in the subsequent service phase, so we
address them as {\em servkey} and {\em servticket} respectively.
The lifetime of a servkey is a few minutes.

Hence, $A$ may start the {\sf service} phase. She sends $B$ a
two-component message including the servticket and an
authenticator sealed by the servkey containing her identity and a
new timestamp $\Ttre$. Upon reception of the message, $B$ decrypts
the servticket, extracts the servkey and checks the validity of
its timestamp $\Ts$. Then, $B$ decrypts the authenticator using
the servkey and checks the validity of $\Ttre$. Finally, $B$
increments $\Ttre$, seals it by the servkey and sends it back to
$A$.
\section{Analysing Kerberos}\label{kerberosanalysis}
\begin{figure}[htbp]
\centering
\psfrag{all_keys --> private}{{\tiny allkeys $\rightarrow private$}}
\psfrag{A}{{\tiny $A$}}
\psfrag{B}{{\tiny $B$}}
\psfrag{KAS}{{\tiny $\kas$}}
\psfrag{TGS}{{\tiny $\tgs$}}
\psfrag{a --> public}{{\tiny $\langle a \rangle \rightarrow public$}}
\psfrag{b --> public}{{\tiny $\langle b \rangle \rightarrow public$}}
\psfrag{tgs --> public}{{\tiny $\langle tgs \rangle \rightarrow public$}}
\psfrag{kas --> public}{{\tiny $\langle kas \rangle \rightarrow public$}}
\psfrag{Ka --> private}{{\tiny $\langle \Ka \rangle \rightarrow private$}}
\psfrag{Kb --> private}{{\tiny $\langle \Kb \rangle \rightarrow private$}}
    \includegraphics[scale=.5]{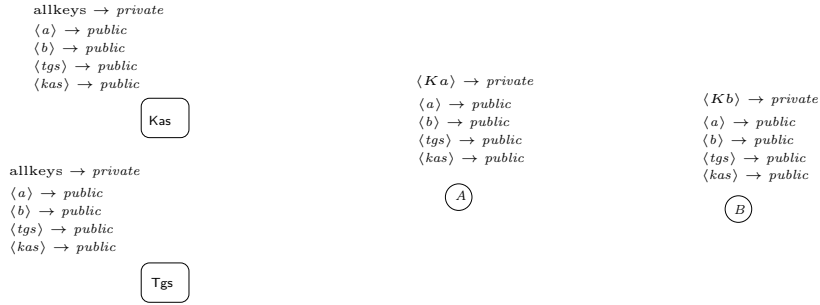}
    \caption{The initial SCSP for Kerberos (fragment).}
    \label{fig:kerberos_initial}
\end{figure}

As a start, we build the initial SCSP for Kerberos.
\RefFig{fig:kerberos_initial} shows the fragment pertaining to
principals $A$ and $B$. The assignment $allkeys \rightarrow private$
signifies that the constraint assigns level $private$ to all principals'
long-term keys.

\begin{figure}[htbp]
\centering
\psfrag{all_keys --> private}{{\tiny allkeys $\rightarrow private$}}
\psfrag{a --> public}{{\tiny $\langle a \rangle \rightarrow public$}}
\psfrag{b --> public}{{\tiny $\langle b \rangle \rightarrow public$}}
\psfrag{tgs --> public}{{\tiny $\langle tgs \rangle \rightarrow public$}}
\psfrag{kas --> public}{{\tiny $\langle kas \rangle \rightarrow public$}}
\psfrag{Ka --> private}{{\tiny $\langle \Ka \rangle \rightarrow private$}}
\psfrag{Kb --> private}{{\tiny $\langle \Kb \rangle \rightarrow private$}}

\psfrag{authTicket = {| a,tgs,authK,Ta |}_Ktgs --> private}{{\tiny
    $\at = \comp{a,tgs,\ak,\Ta}_{\Ktgs}$}}

\psfrag{authenticatorTGS = {| a,T2 |}_authK --> traded1}{{\tiny
    $\authenticatoruno = \comp{a,\Tdue}_{\ak}$}}

\psfrag{servTicket = {| a,b,servK,Ts |}_Kb --> private}{{\tiny
    $\st =\comp{a,b,\sk,\Tt}_{\Kb}$}}

\psfrag{authenticatorB = {| a,T3 |}_servK --> traded3}{{\tiny
    $\authenticatordue = \comp{a,\Ttre}_{\sk}$}}

\psfrag{authenticatorA = {| T3+1 |}_servK --> traded4}{{\tiny
    $\authenticatortre = \comp{\Ttre + 1}_{\sk}$}}

\psfrag{1:<a,tgs,T1> --> public}{{\tiny
    $1:~~\langle a,tgs,\Tuno \rangle \rightarrow public$}}

\psfrag{2:<{| authK,tgs,Ta,authTicket |}_Ka> = -> traded1}{{\tiny
    $2:~~\langle \comp{\ak,tgs,\Ta,\at}_{\Ka}\rangle \rightarrow traded_1$}}

\psfrag{3:<authTicket,authenticatorTGS,B> --> traded2}{{\tiny
    $3:~~\langle \at, \authenticatoruno, b\rangle \rightarrow traded_2$}}

\psfrag{4:<{| servK,B,Ts,servTicket |}_authK> --> traded3}{{\tiny
    $4:~~\langle \comp{\sk,b,\Ts,\st}_{\ak}\rangle \rightarrow traded_3$}}

\psfrag{5:<servTicket,authenticatorB> --> traded4}{{\tiny
    $5:~~\langle \st,\authenticatordue\rangle \rightarrow traded_4$}}

\psfrag{6:authenticatorA --> traded5}{{\tiny
    $6:~~\langle \authenticatortre \rangle \rightarrow traded_5$}}

\psfrag{A}{{\tiny $A$}}
\psfrag{B}{{\tiny $B$}}
\psfrag{KAS}{{\tiny $\kas$}}
\psfrag{TGS}{{\tiny $\tgs$}}

\psfrag{1}{{\tiny $1$}}
\psfrag{2}{{\tiny $2$}}
\psfrag{3}{{\tiny $3$}}
\psfrag{4}{{\tiny $4$}}
\psfrag{5}{{\tiny $5$}}
\psfrag{6}{{\tiny $6$}}

    \includegraphics[scale=.53]{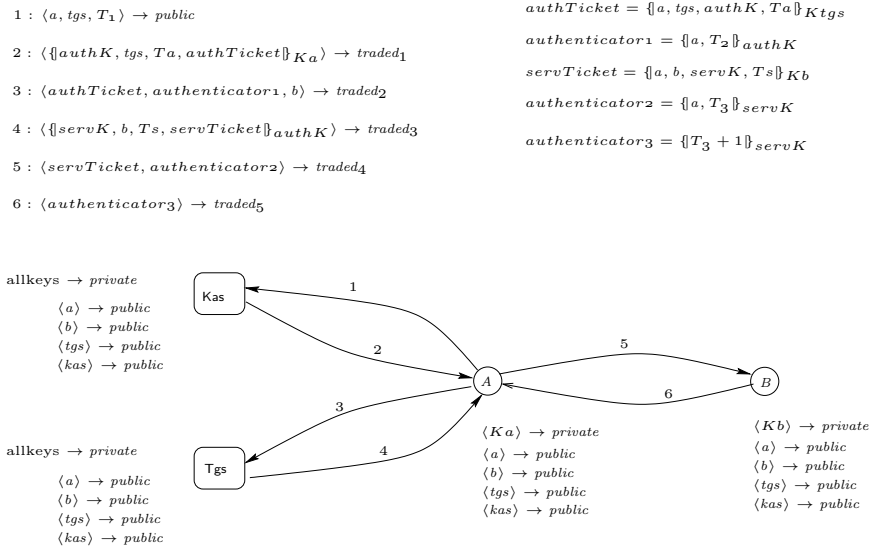}
    \caption{The policy SCSP for Kerberos (fragment).}
    \label{fig:kerberos_policy}
\end{figure}

Then, we build the policy SCSP for Kerberos using algorithm
{\sc BuildPolicySCSP} (\RefFig{procedure}).
\RefFig{fig:kerberos_policy} shows the fragment pertaining to principals
$A$ and $B$. The components that are specific of the session between $A$ and
$B$, such as timestamps and session keys, are not indexed
for simplicity. We remark that the security levels of all other principals
on the authkey $\ak$ and on the servkey $\sk$ are $unknown$.
\subsection{Confidentiality}\label{kerconf}
The preliminary analysis of confidentiality conducted on the policy SCSP
in \RefFig{fig:kerberos_policy} highlights that the late protocol
messages get worse security levels than the initial ones do.
For example, by definition~\ref{deflconf}, there is $traded_3$-confidentiality
of $\sk$ for $B$. By the same definition, it is crucial to observe that
$A$ gets $\ak$ as $traded_1$-confidential, but gets $\sk$ as $traded_3$-confidential.
So, if we consider a potential confidentiality attack whereby
$A$ looses $\ak$ to some malicious principal other than $B$,
and another potential confidentiality attack whereby
$A$ or $B$ loose $\sk$ to some malicious principal,
the former would be a worse confidentiality attack
than the latter, by definition~\ref{pre-conf-att}.
Indeed, having $\ak$ available, one can obtain $\sk$ from decryption
and splitting of message 4.

We also conduct an empirical analysis of confidentiality by considering,
as example a {\em known-ciphertext attack}~\cite{stinsonbook}
mounted by some malicious principal $C$
on the authenticator of message 3 to discover the authkey pertaining to
a principal $A$ (and $\tgs$).
We briefly remind how such an attack works.
Since both principal names and timestamps are public,
$C$ knows the body of the authenticator
with a good approximation --- she should just try out all timestamps
of, say, the last day. First, she invents a key, encrypts the known
body with it, and checks whether the result matches the encrypted
authenticator fetched from the network. If not, $C$ ``refines''
her key~\cite{stinsonbook}
and iterates the procedure until she obtains the same ciphertext
as the authenticator. At this stage, she holds the encryption key,
alias the authkey, because encryption is injective. The entire tampering
took place off line.

Along with the authkey for $A$, principal $C$ also saves a copy of the
corresponding authticket by splitting message 3 into its components.
Then, $C$ forwards
message 3, unaltered, to $\tgs$, so $A$ can continue and terminate
the session accessing some resource $B$.
A glimpse to \RefFig{fig:kerberos_prot} shows that $C$ is now in
a position to conduct, for the lifetime of the authkey,
the {\sf Authorisation} and {\sf Service}
phases while he masquerades as $A$ with some principal $D$.
To do so, $C$ forges an instance 3$'$ of message 3 by using the authticket
just learnt, by refreshing the timestamp inside the authenticator (which he
can do because he knows the authkey), and by mentioning the chosen
principal $D$.
As $\tgs$ believes
that the message comes from $A$, $\tgs$ replies to $A$ with a message 4$'$
containing some fresh servkey
meant for $A$ and $D$.
%
Having intercepted 4$'$, $C$ learns the servkey and therefore can forge
an instance 5$'$ for $D$ of message 5. Finally, $C$ intercepts 6$'$ and
the session terminates without $A$'s participation.

\begin{figure}[htbp]
\centering

\psfrag{authTicket = {| a,tgs,authK,Ta |}_Ktgs --> private}{{\tiny
    $\at = \comp{a,tgs,\ak,\Ta}_{\Ktgs}$}}

\psfrag{authenticator1 = {| a,T2 |}_authK --> traded1}{{\tiny
    $\authenticatoruno = \comp{a,\Tdue}_{\ak}$}}

\psfrag{servTicket = {| a,b,servK,Ts |}_Kb --> private}{{\tiny
    $\st = \comp{a,b,\sk,\Tt}_{\Kb}$}}

\psfrag{servTicket' = {| a,d,servK',Ts' |}_Kd --> private}{{\tiny
    $\st' = \comp{a,d,\sk',\Tt'}_{\Kd}$}}

\psfrag{authenticator1' = {| A,T2' |}_authK --> private}{{\tiny
    $\authenticatoruno'= \comp{a,\Tdue'}_{\ak}$}}

\psfrag{authenticator2' = {| A,T3' |}_servK' --> private}{{\tiny
    $\authenticatordue'= \comp{a,\Ttre'}_{\sk'}$}}

\psfrag{authenticator3' = {| T3'+1|}_servK' --> private}{{\tiny
    $\authenticatortre'= \comp{\Ttre'+1}_{\sk'}$}}

\psfrag{3:<authTicket,authenticator1,B> --> traded2}{{\tiny
    $3:~~\langle \at,\authenticatoruno,b\rangle \rightarrow traded_2$}}

\psfrag{9:<authTicket,authenticator1,B> --> traded3}{{\tiny
     $\bar{3}:~~\langle \at,\authenticatoruno,b\rangle \rightarrow traded_3$}}

\psfrag{4:<{| servK,B,Ts,servTicket |}_authK> --> traded3}{{\tiny
    $4:~~\langle \comp{\sk,b,\Ts,\st}_{\ak}\rangle \rightarrow traded_3$}}

\psfrag{3':<authTicket,authenticator1',D> --> traded1}{{\tiny
    $3':~~\langle \at,\authenticatoruno',d\rangle \rightarrow traded_3$}}

\psfrag{4':<{| servK',D,Ts',servTicket' |}_authK> --> traded3}{{\tiny
    $4':~~\langle \comp{\sk',d,\Ts',\st'}_{\ak}\rangle \rightarrow traded_4$}}

\psfrag{5':<servTicket',authenticator2'>-->traded3}{{\tiny
    $5':~~\langle \st',\authenticatordue' \rangle \rightarrow traded_5$}}

\psfrag{6':<authenticator3'>-->traded3}{{\tiny
    $6':~~\langle \authenticatortre' \rangle \rightarrow traded_6$}}

\psfrag{authK --> private}{{\tiny $\langle \ak \rangle \rightarrow private$}}

\psfrag{authTicket-->}{{\tiny $\langle \at \rangle \rightarrow traded_2$}}

\psfrag{A}{{\tiny $A$}}
\psfrag{C}{{\tiny $C$}}
\psfrag{D}{{\tiny $D$}}
\psfrag{TGS}{{\tiny $\tgs$}}

\psfrag{3}{{\tiny $3$}}
\psfrag{9}{{\tiny $\bar{3}$}}
\psfrag{4}{{\tiny $4$}}
\psfrag{3'}{{\tiny $3'$}}
\psfrag{4'}{{\tiny $4'$}}
\psfrag{5'}{{\tiny $5'$}}
\psfrag{6'}{{\tiny $6'$}}

    \includegraphics[scale=.53]{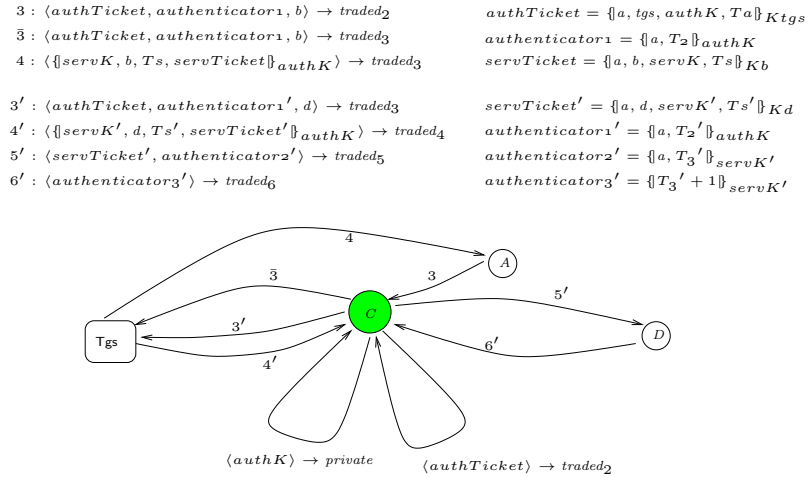}
    \caption{An imputable SCSP for Kerberos (fragment).}
    \label{fig:bruteforce2}

\Commento{
C ottiene authK privato per cryptanalisi.
C ottiene authticket per splitting da 3, quindi traded2.
C rimbalza 3 come traded3,
  quindi TGS si registra authK come traded3.
  (mentre nel policy se lo registrava traded2 perche' C non aveva
   interferito).
E la sessione fra A e B si completa.

C riceve 4' come traded4, e quindi
servticket' come traded4 e servK' come traded4.
Quindi 5' diventa traded4, spedito traded5.

D ottiene servK' come traded5.
Decriptando authenticator2', D ottiene T3' come traded5.
Crea T3'+1 che resta traded5 (omettere).
Crea authenticator3' come traded5, e lo spedisce come traded6.

Confidentiality atttack su authK da C.
                        su authK da TGS.
}

\end{figure}

Our algorithm {\sc BuildImputableSCSP}
executed on the network configuration
just described produces the imputable SCSP in
\RefFig{fig:bruteforce2}.
The SCSP omits the constraint corresponding to the {\sf Authentication} phase
between $A$ and $\kas$.
Because $C$ intercepts message 3, constraint 3 is stated between $A$ and $C$.
Projecting that constraint on $C$, we have that $C$'s security level on
message $\at,\authenticatoruno,b$ is $traded_2$.
By splitting this message, $C$ discovers the authticket,
so the entailment relation
states a unary constraint on $C$ assigning $traded_2$ to $\at$.
Another unary constraint on $C$ assigns $private$ to $\ak$, which is found by
cryptanalysis.

Constraint $\bar{3}$ between $C$ and $\tgs$ assigns $traded_3$ to message
3 because of $C$'s rerouting.
Projecting that constraint on $\tgs$, we have by entailment
that $\tgs$'s security level on $\ak$ goes down to $traded_3$, whereas
it was $traded_2$ in the policy SCSP.
Constraint 4 formalises $\tgs$'s reply to
$A$, while the constraints for the rest of the session between $A$ and $B$ are
omitted. Constraints 3$'$, 4$'$, 5$'$, and 6$'$ formalise the session between
$C$, $\tgs$, and $D$.

At this stage, we can conduct an empirical analysis of confidentiality for
each of the agents involved in this imputable SCSP.
By definition~\ref{deflconf},
$\at$, $\authenticatoruno'$, and $\ak$ are each
$traded_3$-confidential for $\tgs$
in this problem. Since they were $traded_2$-confidential in the policy SCSP,
we conclude by definition~\ref{emp-conf-att}
that there is a confidentiality attack by $\tgs$ on each of these
messages in the imputable SCSP considered here.
The attacks signal $C$'s manipulation of messsage 3.

The imputable SCSP also achieves $private$-confidentiality
of $\ak$ for $C$, whereas the policy SCSP achieved $unknown$-confidentiality
of $\ak$ for $C$. Therefore,
there is a confidentiality attack by $C$ on $\ak$ in this SCSP.
Likewise, there is a confidentiality attack by $C$ on $\at$.
From constraint 4$'$ we have by entailment that $C$'s security level on
$\st'$ and on $\sk'$ is $traded_4$ rather than $unknown$ as in the policy SCSP,
hence we find other confidentiality attacks by $C$ on each of these messages.

There are also confidentiality attacks by $D$,
who gets $\st'$, $\authenticatordue'$, and
$\sk'$ as $traded_5$, rather than $traded_4$.
\subsection{Authentication}\label{kerauth}
We now focus on the fragment of policy SCSP for Kerberos given in
\RefFig{fig:kerberos_policy} to conduct the preliminary analysis of the
authentication goal.

By definition~\ref{deflauth}, there is
$traded_2$-authentication of $A$ with $\tgs$ in the policy SCSP.
The definition holds for message 3, whose first two components speak about $A$.
Also, there is
$traded_4$-authentication of $A$ with $B$ thanks to message 5, and
$traded_5$-authentication of $B$ with $A$ due to message 6.
While it is obvious that message 5 speaks about $A$, it is less obvious that
message 6 speaks about $B$. This is due to the use of
a servkey that is associated to $B$.

We observe that authentication of $B$ with $A$ is weaker than authentication
of $A$ with $B$ even in the ideal conditions formalised by the policy SCSP.
Intuitively, this is due to the fact that the servkey has been handled both
by $A$ and $B$ rather than just by $A$.
Hence, by definition~\ref{pre-auth-att}, a principal $C$'s masquerading as
$A$ with $B$ would be a worse authentication attack than a principal $D$'s
masquerading as $B$ with $A$.

An empirical analysis of authentication can be conducted
on the imputable SCSP in \RefFig{fig:bruteforce2}. That SCSP achieves
$traded_5$-authentication of $A$ with $B$ thanks to message 5, and
$traded_6$-authentication of $B$ with $A$ due to message 6.
Comparing these properties with the equivalent ones holding in the policy
SCSP, which we have seen above, we can conclude
by definition~\ref{emp-auth-att} that the imputable SCSP considered
hides an authentication attack on $B$ by means of $A$, and an authentication
attack on $A$ by means of $B$. They are due to $C$'s interception of message
3, which has lowered the legitimate protocol participants' security levels
on the subsequent messages.

It is important to emphasize that these authentication attacks could not
be captured by an equivalent definition of authentication based on crisp,
rather than soft, constraints. The definition in fact holds in the policy
SCSP as well as in the imputable SCSP.
What differentiates the two SCSPs is merely the security level
characterising the goal.

\section{Conclusions}\label{sec:concl}
We have developed a new framework for analysing security
protocols, based on a recent
kernel~\cite{security-padl01,bella-bista-cambridge02}. Soft constraint
programming allows us to conduct a fine analysis of the
confidentiality and authentication goals that a protocol attempts
to achieve. Using the security levels, we can formally claim that
a configuration induced by a protocol achieves a certain level of
confidentiality or authentication. That configuration may be ideal
if every principal behaves according to the protocol, as
formalised by the policy SCSP; or, it may arise from the protocol
execution in the real world, where some principal may have acted
maliciously, as formalised by an imputable SCSP.
We can formally express that different principals
participating in the same protocol session obtain different forms
of those goals. We might even compare the forms of the same goal
as achieved by different protocols.

Our new threat model where each principal is a potential attacker
working for his own sake has allowed us to detect a novel
attack on the asymmetric
Needham-Schroeder protocol. Once $C$ masquerades as $A$ with $B$,
agent $B$ indeliberately gets hold of a nonce that was not meant for him.
At this stage, $B$ might decide to exploit this extra knowledge, and begin
to act maliciously. Our imputable SCSP modelling the scenario reveals
that $B$'s security level on the nonce is lower than that
allowed by the policy.

There is some work related to our analysis of Kerberos, such as
Mitchell et al.'s analysis by model checking~\cite{mitchell}.
They consider
a version of Kerberos simplified of timestamps and
lifetimes --- hence authkeys and servkeys cannot be distinguished ---
and establish that a small system with an initiator, a responder,
$\kas$ and $\tgs$ keeps the two session keys secure from the spy.
Bella and
Paulson~\cite{mioesorics} verify by theorem proving a version
with timestamps of the same protocol. They do prove that using a lost
authkey will let the spy obtain a servkey. On top of this, one can
informally
deduce that the first key is more important than the second in
terms of confidentiality. By contrast, our preliminary analysis of the
protocol states formally
that the authkey is $traded_1$-confidential and the servkey
is $traded_3$-confidential (\S\ref{kerconf}). Another finding is
the difference between authentication of initiator with
responder and vice versa (\S\ref{kerauth}).

Some recent research exists that is loosely related to ours.
Millen and Shamatikov \cite{millen-constraint} map the existence
of a {\em strand} representing the attack upon a constraint
problem. Comon { \em et al.} \cite{ping-pong},  and Amadio and
Charatonik \cite{amadio-setconstraint} solve confidentiality and
reachability using {\em Set-Based Constraint}
\cite{pacholski97set}. By constrast, we build suitable constraint
problems for the analysis of a global network configuration where
any principals (not just one) can behave maliciously. In doing so,
we also analyse the safety of the system in terms of the
consequences of a deliberate attack on the environment.
The idea of {\em refinements}~\cite{refinements} is also somewhat related to
our use of levels. In that case the original protocol
must be specialised in order to be able to map the known/unknown level
over the set of levels specified by the policy.
The policy have also to specify how the levels have to be changed w.r.t.
each operation described in the protocol.
Abstract interpretation techniques
(much in the spirit of those used by Bistarelli et al.~\cite{CSPabstraction})
can be used as a next step to deal with unbounded
participants/sessions/messages.

While mechanical analysis was outside our aims,
we have implementated a mechanical checker for $l$-confidentiality on top of
the existing {\em Constraint Handling Rule} (CHR) framework~\cite{sac2002}.
For example, when we input the policy SCSP for the Needham-Schroeder protocol
and the imputable SCSP corresponding to Lowe's attack, the checker outputs
\begin{verbatim}
checking(agent(a))
checking(agent(b))
   attack(n_a, policy_level(unknown), attack_level(traded_1))
checking(agent(c))
   attack(enk(k(a),pair(n_a,n_b)), policy_level(unknown),
                                   attack_level(traded_1))
   attack(n_b, policy_level(unknown), attack_level(traded1))
\end{verbatim}

The syntax seems to be self-explanatory.
Line two reveals the new attack we have found on $B$, who has lowered
his security level on $\Na$ from $unknown$ to $traded_1$.
Likewise, line three denounces that
not only has $C$ got hold of the nonce $\Nb$ but also of the message
$\comp{\Na,\Nb}_\Ka$ (which was meant for $A$ and not for $B$)
that contains it.

At this stage, integrating our framework with model-checking tools
appears to be a straightforward exercise.
The entailment relation must be extended by a rule per each of the
protocol messages in order to compute their security levels.
Hence, our constraints would be upgraded much the way
multisets are rewritten in the work by Cervesato et
al.~\cite{cervesato99metanotation} (though they only focus on a single
attacker and their properties are classical yes/no properties).
Then, once suitable size limits are stated,
the imputable SCSPs could be exhaustively
generated and checked against our definitions of
confidentiality and authentication.
Alternatively, the protocol
verifier might use our framework for a finer analysis of network
configurations generated using other techniques.

\paragraph{Acknowledgements.} We are indebted to Martin Abadi for
invaluable suggestions, and to Michael Marte and Thom Fruewirth for the
preliminary implementation of the checker. Discussions with Luca Vigan\`o,
Nancy Durgin and Fabio Massacci were important.
Comments by the anonymous referees helped us improve the presentation
considerably.

\bibliographystyle{acmtrans}
\bibliography{acronyms,securitybiblio}
\label{lastpage}
\end{document}